\documentclass[twocolumn,preprintnumbers,amsmath,amssymb]{revtex4}
\usepackage{graphicx} 
\usepackage{setspace} 
\usepackage{dcolumn} 
\usepackage{bm} 
\usepackage{natbib}
\usepackage{hyperref}
\hypersetup{colorlinks=true,linkcolor=blue,citecolor=blue,urlcolor=blue}
\usepackage{epsfig}
\usepackage{amsmath}
\begin{document}
\title{Thermoelectric properties of high-entropy wolframite oxide: (CoCuNiFeZn)$_{1-x}$Ga$_x$WO$_4$}
\author{Ashutosh Kumar}
\author{Adrien Moll\footnote{Email: adrien.moll@universite-paris-saclay.fr}}
\author{Mouhamad Navid Mouhamadsiradjoudine}
\author{Francois Brisset}
\author{David Bérardan}
\author{Nita Dragoe\footnote{Email: nita.dragoe@universite-paris-saclay.fr}}

\affiliation{ICMMO (UMR CNRS 8182), Université Paris-Saclay, F-91405 Orsay, France}
\date{\today}
\begin{abstract}
In this report, the synthesis of high-entropy wolframite oxide (CoCuNiFeZn)$_{1-x}$Ga$_x$WO$_4$ through standard solid-state route followed by spark plasma sintering (SPS) and their structural, microstructural, and thermoelectric properties are investigated. X-ray diffraction pattern followed by pattern matching refinement shows monoclinic structure with volume of the unit cell decreasing with increasing Ga content. The optical band gap for these oxides shows a cocktail effect in high entropy configuration. The Seebeck coefficient indicates electrons as dominating charge carriers with a non-degenerate behavior. The electrical resistivity decreases with increasing temperature depicting a semiconducting nature. Thermal conductivity in high-entropy samples ($\kappa\sim$2.1 W/m·K $@$ 300\,K) is significantly lower as compared to MgWO$_4$ ($\kappa\sim$11.5 W/m·K $@$ 300\,K), which can be explained by the strong phonon scattering due to large lattice disorder in high entropy configuration. The thermoelectric figure of merit zT increases with Ga doping via modifying all three thermoelectric parameters positively. 
\end{abstract}
\maketitle
\section{Introduction}
In recent years, the field of high-entropy functional materials has grown exponentially, due to the potential of these new materials for various applications. Indeed, populating a crystallographic site with five or more equimolar cations can lead to many novel high-entropy and/or entropy-stabilized phases having promising electronic, magnetic, or thermal properties.\cite{sarkar2021magnetic, musico2020emergent, dragoe2019order, kumar2023magnetic, mazza2021charge, kumar2023novel, rost2015entropy} For example, these include long-range antiferromagnetic ordering,\cite{zhang2019long} colossal dielectric constants,\cite{berardan2016colossal} superionic conductivity,\cite{berardan2016room} substantial lithium storage capacity\cite{sarkar2018high} making them suitable as cathode materials in lithium-ion batteries,\cite{lun2021cation} protonic conductivities,\cite{gazda2020novel} etc. Among these functional properties, high entropy materials could be interesting for thermoelectric applications due to their random distribution of several cations at a single-crystallographic site as shown in some recent studies.\cite{kumar2023thermoelectric, banerjee2020high}\\
A good thermoelectric (TE) material requires a large Seebeck coefficient ($\alpha$), small electrical resistivity ($\rho$), and low thermal conductivity ($\kappa$) to achieve a high figure of merit as zT= $\alpha^2$T/($\rho$·$\kappa$), where T is the absolute temperature. Strong correlation between these thermoelectric parameters often limits the value of zT in a family of materials, which could be significantly improved by decoupling electron and phonon transport. Oxide materials usually have better chemical and thermal stability at high temperature as compared to state-of-the-art TE materials, however their efficiency is generally poor due to lower electrical conductivity and higher thermal conductivity. Several approaches have been adopted in the past to improve the zT in oxide materials that includes doping/substitutions,\cite{jood2011doped, kumar2018improvement} composite methods,\cite{chen2015one, kumar2022effect, kumar2020graphene} nanostructuring,\cite{koumoto2010oxide, kumar2021improved} phonon glass electron crystal approach,\cite{daniels2017phonon} etc. High entropy materials have shown promise in recent time to decouple electron and phonon transport owing to the presence of multiple-cations at a single crystallographic site that may result in the (i) suppression of $\kappa_{ph}$ due to large lattice distortion and mass fluctuation, (ii) enlarged $\alpha$ due to possible band degeneracy,\cite{jiang2021entropy} and (iii) tuned-electrical conductivity as heavy-doping may be viable in the high-entropy materials. In recent times, the high entropy approach has been effective in optimizing TE parameters, especially in designing phonon thermal conductivity ($\kappa_{ph}$). A significant reduction in $\kappa_{ph}$ has been reported in high-entropy SrTiO$_3$ perovskites, where both Sr and Ti site has been populated with five or more cations and their influence on $\kappa_{ph}$ has been discussed. Zhang et al., demonstrated a strong reduction ($\sim$65\%) in $\kappa_{ph}$ in (Sr$_{0.25}$Ca$_{0.25}$Ba$_{0.25}$RE$_{0.25}$)TiO$_3$ with different rare earth (elements) as compared to SrTiO$_3$ at 300\,K\cite{zhang2023high}, $\kappa$=0.7 W/m·K at 1100\,K is observed for Sr(Ti$_{0.2}$Fe$_{0.2}$Mo$_{0.2}$Nb$_{0.2}$Cr$_{0.2}$)O$_3$.\cite{banerjee2020high}Yang et al. improved the TE properties of (Ca$_{0.35}$Sr$_{0.2}$Ba$_{0.15}$Na$_{0.2}$Bi$_{0.1}$)$_3$Co$_4$O$_9$ via entropy engineering at Ca site that simultaneously optimize $\kappa_{ph}$ and $\sigma$ resulting in a maximum zT of 0.3 at 973\,K, which is around 2.5 times higher than Ca$_3$Co$_4$O$_9$.\cite{yang2023ultralow} Kumar et al., showed a maximum zT of 0.23 at 350\,K for high entropy rare-earth cobaltates, which is one of the highest values of zT near room temperature for polycrystalline oxide materials.\cite{kumar2023thermoelectric} These results indicate that the random distribution of cations results in significant scattering of phonons that results in reduced $\kappa_{ph}$, beneficial for TE properties.\\
The high entropy tungstate (MnFeCoNiCuZn)WO$_4$ has been proposed very recently and showed interest in magnetic, optical and electrocatalytic properties.\cite{katzbaer2023synthesis} Herein, we present a series of high entropy wolframite oxide with Ga-doping (CoCuNiFeZn)$_{1-x}$Ga$_x$WO$_4$ (0 $\leq$ \textit{x} $\leq$ 0.1) and discuss their structural, microstructural, optical and thermoelectric properties.\\
\section{Experimental Section}
High-entropy (CoCuNiFeZn)$_{1-x}$Ga$_x$WO$_4$ (0$\leq$\textit{x}$\leq$0.1) oxides were synthesized using standard solid-state reaction methods. Stoichiometric amount of Co$_3$O$_4$ (Alfa Aesar 99.7\%), CuO (Alfa Aesar 99.7\%), NiO (Johnson Matthey, 99.9\%), Fe$_2$O$_3$ (Alfa Aesar 99.99\%), ZnO (Alfa Aesar 99.99\%), Ga$_2$O$_3$ (Alfa Aesar 99.99\%), and WO$_3$ (Schuchard München 99.97\%) were mixed using planetary ball milling (Fritsch Pulverisette 7 Premium Line) at 350 rpm for 20 cycles (5 min on and 2 min off time) in agate jars of 20 mL and 5 balls with 10\,mm of diameter. The obtained mixture was pressed in the form of pellets and the reaction was performed at 1173\,K for 24 hours in air followed by quenching. The pellets were crushed into fine powder followed by ball milling. These fine powders were consolidated with spark plasma sintering using a Dr. Sinter 515S Syntex setup at 1123\,K for 5 minutes with a heating and cooling rate of 100\,K/min under 100\,MPa and using graphite molds. Due to possible carbon contamination and reduction of the materials during SPS, the sintered pellets were annealed in air at 1173\,K for 24 hours and quenched.The relative densities were measured by Archimed's method and were $>$95\%. The samples were cut to proper dimensions for thermoelectric measurement (Seebeck coefficient, electrical resistivity and thermal conductivity). The structural characterization of the samples was performed by X-ray diffraction on powders using a Bruker D8 Advance X-ray diffractometer (Cu K$_{\alpha}$ radiations). Surface morphology and chemical compositions were confirmed using scanning electron microscopy (SEM-FEG Zeiss Sigma HD) equipped with energy-dispersive X-ray spectroscopy (EDS). The optical band gap was obtained from UV-Visible-NIR spectroscopy measurement using Agilent Cary 5000 UV-Vis-NIR spectrometer in diffuse reflective spectra (DRS) mode.
The estimation of band gap is based on the Tauc method, based on the assumption of energy-dependent absorption coefficient ($a$)  $(a.h\nu)^{1/n}=A(h\nu-E_g)$, where $h$ is Planck constant, $\nu$ is the photon's frequency, E$_g$ is the band gap, and $A$ is a constant. The value of $n$ is considered as 1/2 and 2 for direct and indirect transition, respectively. In order to determine the band gap in DRS mode, Kubelka and Munk suggested a method to transform the measured reflectance spectra to the corresponding absorption spectra by applying the Kubelka-Munk function ($F(r)$), defined as $F(r)=(1-R)^{2}/2R$, where $R$ is the reflectance of an infinitely thick specimen.\cite{makula2018correctly} Therefore, the above equation simplifies to $(F(r).h\nu)^{(1/n)}=A(h\nu-E_g)$. The electrical resistivity and Seebeck coefficient were measured using homemade instruments over a wide temperature range of 300\,K-1150\,K under air in a standard four-probe configuration.\cite{byl2012experimental} The transport measurements were performed in both heating and cooling modes and the results are identical. However, for clarity of presentation, only measurement data during heating are presented. The thermal conductivity ($\kappa$) of the samples was calculated using $\kappa=D·\rho_s·C_p$, where the thermal diffusivity ($D$) was measured using the laser flash technique (Netzsch LFA 427), and sample density ($\rho_s$) was measured by Archimed’s method. The specific heat capacity ($C_p$) was calculated using the Dulong-Petits law.\\
\section{Results and Discussion} 
\begin{figure*}
\centering
  \includegraphics[width=0.95\linewidth]{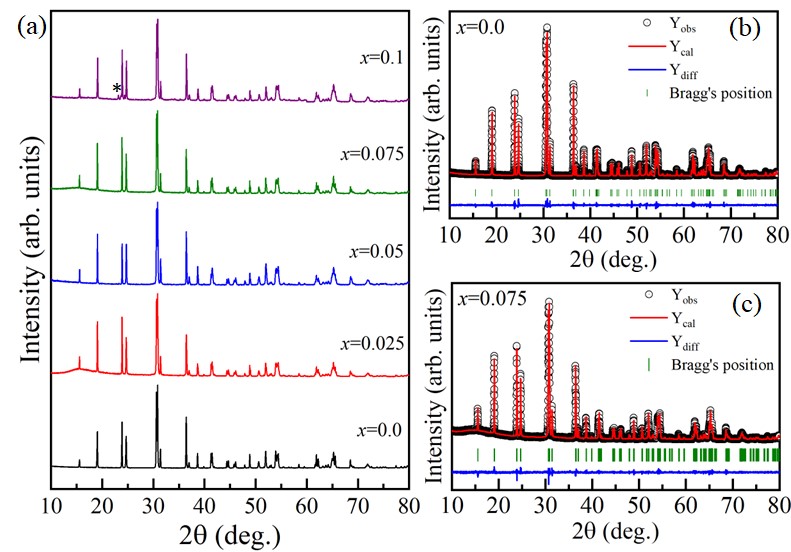}
  \caption{(a) X-ray diffraction (XRD) patterns for (CoCuNiFeZn)$_{1-x}$Ga$_x$WO$_4$ (0$\leq$\textit{x}$\leq$0.1), Rietveld refinement patterns for (b) \textit{x}=0.0, (c) \textit{x}=0.075.}
  \label{fig1a}
\end{figure*}

\begin{figure*}
\centering
  \includegraphics[width=0.8\linewidth]{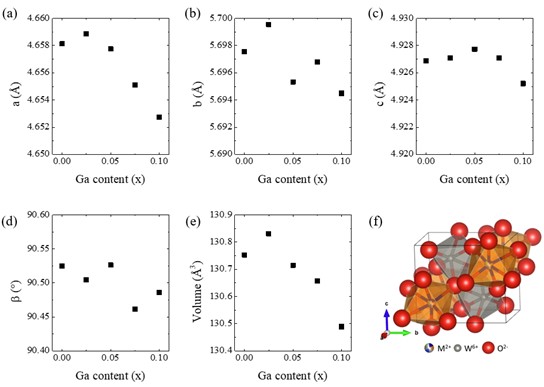}
  \caption{Lattice parameters (a, b, c), $\beta$, volume of unit cell (V) obtained from pattern matching refinement and representation of the unit cell for (CoCuNiFeZn)$_{1-x}$Ga$_x$WO$_4$ (0$\leq$\textit{x}$\leq$0.1).}
  \label{fig2a}
\end{figure*}

\begin{figure}
\centering
  \includegraphics[width=0.99\linewidth]{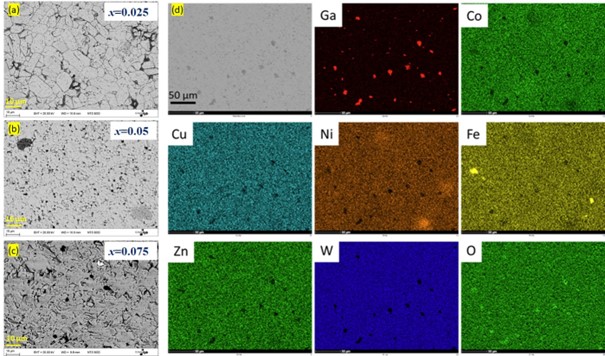}
  \caption{Scanning electron microscopy (SEM) images in back-scattered mode is shown for (CoCuNiFeZn)$_{1-x}$Ga$_x$WO$_4$ (a) \textit{x}=0.025, (b) \textit{x}=0.05, (c) \textit{x}=0.075, (d) EDS mapping of (CoCuNiFeZn)$_{0.95}$Ga$_{0.05}$WO$_4$, i.e., \textit{x}=0.05.}
  \label{fig3a}
\end{figure}

\begin{figure}
\centering
  \includegraphics[width=0.99\linewidth]{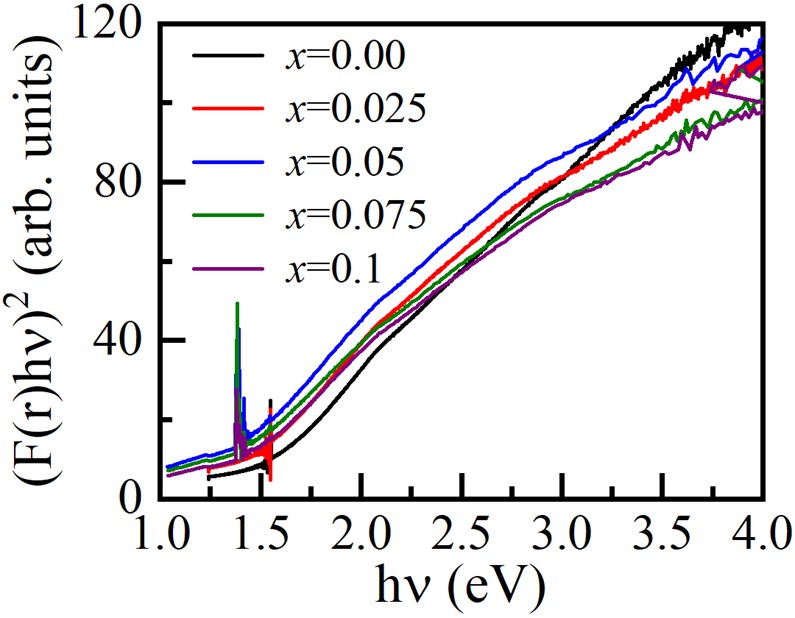}
  \caption{Optical band gap from UV-Vis-NIR spectroscopy measurement for (CoCuNiFeZn)$_{1-x}$Ga$_x$WO$_4$ (0$\leq$\textit{x}$\leq$0.1).}
  \label{fig4a}
\end{figure}

\begin{figure}
\centering
  \includegraphics[width=0.99\linewidth]{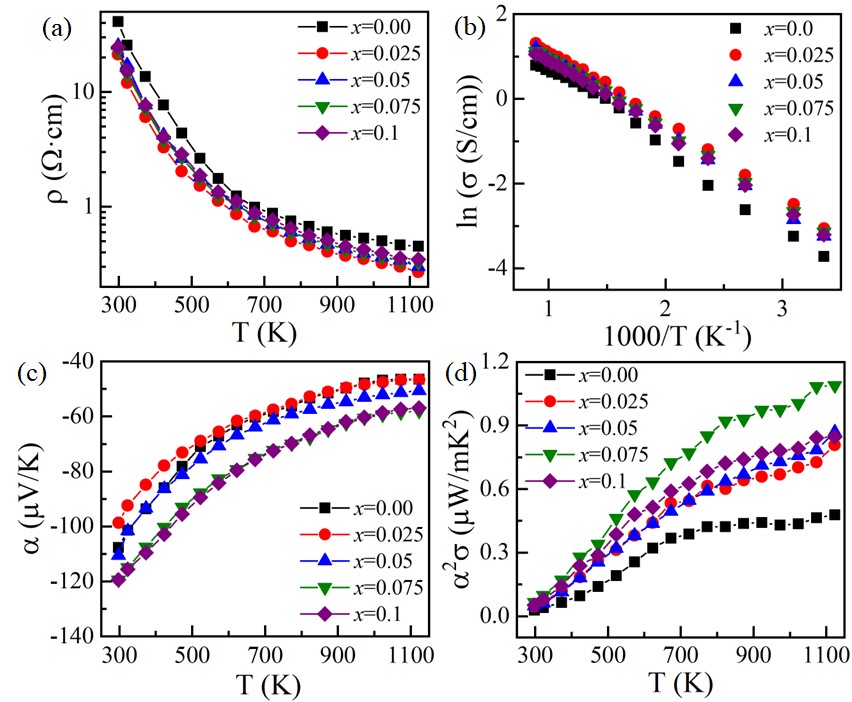}
  \caption{(a) Electrical resistivity ($\rho$), (b) Arrhenius plot: ln($\rho$) vs 1000/T, (c) Seebeck coefficient ($\alpha$), and (d) powder factor ($\alpha^2\sigma$) as a function of temperature for (CoCuNiFeZn)$_{1-x}$Ga$_x$WO$_4$ (0$\leq$\textit{x}$\leq$0.1).}
  \label{fig5a}
\end{figure}

X-ray diffraction patterns for (CoCuNiFeZn)$_{1-x}$Ga$_x$WO$_4$ (0$\leq$\textit{x}$\leq$0.1) are shown in Fig.~\ref{fig1a}(a). The XRD pattern for each sample is indexed with the wolframite structure (space group: \textit{P2/c}, no. 13, PDF 96-101-0643), confirming the formation of the desired material. The samples with Ga doping (\textit{x}) at M$^{2+}$ site show similar structure. Some WO3 (PDF 01-072-1465) impurity traces are only observed mostly at 2$\theta$ = 23.1$^{\circ}$, 23.6$^{\circ}$ and 24.4$^{\circ}$ with \textit{x}=0.1, suggesting a solubility limit. Pattern matching is performed using Fullprof software to confirm the crystalline structure and calculate the lattice parameters. The refined patterns for two compositions (CoCuNiFeZn)$_{1-x}$Ga$_x$WO$_4$ with \textit{x}=0.0 and 0.075 are shown in Fig.~\ref{fig1a}(b-c). The experimental XRD pattern is in good agreement with the calculated one, depicting a single monoclinic phase for (CoCuNiFeZn)$_{1-x}$Ga$_x$WO$_4$ with \textit{x} up to 0.075. The evolution of the lattice parameters and $\beta$ angle with Ga doping, obtained from the Rietveld refinement, is given in Fig.~\ref{fig2a}(a-d). For \textit{x}=0 the lattice parameters are a=4.658$\AA$, b=5.698$\AA$, c=4.927$\AA$, $\beta$=90.53$^{\circ}$. The volume of the unit cell decreases with increasing \textit{x}, as shown in Fig.~\ref{fig2a}(e). A non-monotonic decrease of the lattice parameters is observed when Ga is added, especially for a parameter. This decrease in lattice parameters and volume of the unit cell may be attributed to the reduced ionic radius of Ga as compared to other five metal cations,\cite{shannon1976revised} and confirms that the Ga is introduced into the HEOx structure. However, further experiments are required to get a better understanding of this non-monotonic evolution. The crystal structure is shown in Fig.~\ref{fig2a}(f). The wolframite structure consists of distorted octahedra.\cite{nocerino2023magnetic} Both WO$_6$ and MO$_6$ octahedra arrange themselves into edge-sharing zigzag chains aligned parallel to the c-axis. These two chains are linked to each other through corner sharing. The thermal stability of the sample \textit{x}=0 was tested by annealing at 673\,K for 4 weeks. The material was not decomposed after this heat treatment, suggesting that this material is a high entropy oxide but not an entropy stabilised one.\\
Scanning electron microscopy (SEM) images and EDS mapping of one of the spark plasma sintered samples are shown in Fig.~\ref{fig3a}(a-d) (EDS mapping of other samples are given in SI, Fig. S1, Fig. S2). The compact surface morphology is seen for all the samples with average crystallite size ranging in few micrometers with some porosity consistent with the geometrical density >95\%. However, the samples are not perfectly homogeneous. Although no Ga-containing secondary phases were observed in the XRD patterns, part of the added gallium forms gallium-oxide clusters of about 10 $\mu$m observed in black in Fig.~\ref{fig3a}(b) and in red in the EDS map of Ga in Fig.~\ref{fig3a} and Fig. S1-S2. Besides, the amount of these gallium oxide clusters increases when \textit{x} increases. Nevertheless, the evolution of the lattice parameters observed by XRD (Fig.~\ref{fig2a}) coupled to the presence of Ga within the matrix observed in the EDS mappings confirm that a part of Ga is actually inserted into the HEOx wolframite structure, but in thermodynamic equilibrium with gallium-oxide clusters. Therefore, the Ga content (\textit{x}) corresponds to the nominal composition of the sample but does not depict the real doping level of the HEOx. This equilibrium between Ga inserted in the wolframite structure and Ga segregated in gallium oxide clusters, as well as the possible evolution of the oxygen stoichiometry with Ga$^{3+}$ introduction in MWO$_4$, could possibly be responsible for the non-monotonic evolution of the lattice parameters.\\  
The optical band gap for (CoCuNiFeZn)$_{1-x}$Ga$_x$WO$_4$ (0$\leq$\textit{x}$\leq$0.1) has been obtained from the room-temperature UV-Visible-NIR spectroscopy measurements. The $(F(r)h\nu)^2 vs. h\nu$ curves are shown in Fig.~\ref{fig4a} for all the samples. Surprisingly, the optical band gap for (CoCuNiFeZn)WO$_4$ (\textit{x}=0) is close to 1.6\,eV which is significantly lower than all individual single cation A-site wolframite oxides (from 2.3\,eV for CuWO$_4$, to 4.4\,eV for ZnWO$_4$\cite{ruiz2012pressure, bhuyan2017experimental, lacomba2008optical} We also measured the optical band of 3.36\,eV for MgWO$_4$ in good agreement with these references (see Fig. S3 in SI). Therefore, the mixing of the five cations on the A-site has a large unexpected impact on the optical properties of the materials. Besides, with Ga doping (\textit{x}) at A-site, the optical band gap reduces in (CoCuNiFeZn)WO$_4$ and reaches a minimum value of 1.3\,eV for \textit{x}= 0.05, which further confirms the introduction of Ga in the wolframite structure. Theoretical calculations would be useful to get a better understanding of the optical properties of the high-entropy wolframite samples, which seem to exhibit a cocktail effect.\\
Electrical resistivity ($\rho$) as a function of temperature for (CoCuNiFeZn)$_{1-x}$Ga$_x$WO$_4$ (0$\leq$\textit{x}$\leq$0.1) is shown in Fig.~\ref{fig5a}(a). The decrease in $\rho$ with increasing temperature (d$\rho$/dT$<$0) for all the HEOx samples indicates a semiconducting behavior. For comparison with single-cation wolframites, we measured the resistivity of MgWO$_4$ which was insulating. Literature shows $\rho\sim$10$^3$ $\Omega$.cm for polycrystalline FeWO$_4$ at 300\,K.\cite{schmidbauer1991electrical},   $\rho$~10$^7$ $\Omega$.cm for polycrystalline CuWO$_4$, CoWO$_4$ single crystals, and NiWO$_4$ single crystals at 300\,K\cite{bharati1983electrical, bharati1981electrical, bharati1980electrical}, and $\rho\sim$10$^8$ $\Omega$.cm for polycrystalline ZnWO$_4$ at 570\,K.\cite{dkhilalli2018structural} The large decrease of electrical resistivity compared to single-cation wolframites (at least two orders of magnitude) may be explained by the local chemical and structural disorder in high entropy configuration that may decrease the energy of formation of defects, and so increase the charge carrier concentration. $\rho$ first decreases with Ga doping for \textit{x}=0.025, however it increases with larger doping level, which may be explained by the presence of insulating gallium oxides inclusions in the system. The Arrhenius plot (ln$\rho$ vs. 1000/T) for these HEOx samples is shown in Fig.~\ref{fig5a}(b). The activation energy obtained between 300 K-650 K for these samples varies between 0.17 eV for \textit{x}=0.0 to 0.16 eV for \textit{x}=0.05, which means that charge carriers do not originate from thermal excitations through the bandgap but from the activation of defects. A slight slope change is observed in the Arrhenius plot above about 650\,K, which may be correlated to a spin-state transition of cobalt.\cite{mahendiran1996magnetoresistance} or to a transition from a polaron hopping mechanism close to room temperature to a band-conduction mechanism at high temperature.\cite{bharati1981electrical}\\

Temperature-dependent Seebeck coefficients ($\alpha$) for (CoCuNiFeZn)$_{1-x}$Ga$_x$WO$_4$ (0$\leq$\textit{x}$\leq$0.1) is shown in Fig.~\ref{fig5a}(c). A negative value of $\alpha$ is observed for all the samples, indicating dominating n-type conduction in the system. At 300\,K, $\alpha$ for (CoCuNiFeZn)WO4 is -109 $\mu$V/K and it first decreases (in absolute value) (-100 $\mu$V/K for x=0.025) and then increases (-112 $\mu$V/K for \textit{x}=0.05, -120 $\mu$V/K for \textit{x}=0.075). $\alpha$ decreases with increasing temperature, which confirms the semiconducting behavior of the samples. The power factor ($\alpha^2\sigma$) calculated using $\alpha$ and $\sigma$(=1/$\rho$) is shown in Fig.~\ref{fig5a}(d). The $\alpha^2\sigma$ is enhanced with the introduction of Ga in the structure, and reaches its maximum value for \textit{x}=0.075. However, even if $\rho$ decreases significantly when the temperature increases from 300\,K to 1100\,K, the simultaneous decrease of $\alpha$ leads to low values of $\alpha^2\sigma$ that does not exceed 1 $\mu$W/m·K².\\

\begin{figure}
\centering
  \includegraphics[width=0.99\linewidth]{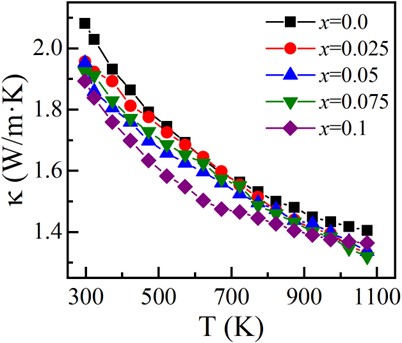}
  \caption{Total thermal conductivity ($\kappa$) as a function of temperature for (CoCuNiFeZn)$_{1-x}$Ga$_x$WO$_4$ (0$\leq$\textit{x}$\leq$0.1).}
  \label{fig6a}
\end{figure}
Total thermal conductivity ($\kappa$) as a function of temperature is shown in Fig.~\ref{fig6a}. $\kappa$ decreases with increasing the temperature for all samples. As seen from the $\rho$(T) measurement, the rise in $\sigma$ (=1/$\rho$) is not significant enough to perturb $\kappa$. The maximum electronic contribution, $\kappa_e$ calculated for these sample is 0.007 W/m·K, which does not perturb $\kappa$. Therefore, the change in $\kappa$ is basically a change in $\kappa_{ph}$. It is noted that the $\kappa$ for the HEOx sample is significantly lower than that of the several single cations M$^{2+}$WO$_4$. For example: simple MgWO4 sample, $\kappa$ is 11.5 W/m·K at 300 K [see Fig. S4], $\kappa$=4.76 @ 300\,K for ZnWO$_4$ (single crystal)\cite{popov2016thermal}, $\kappa$=4.69 @300 K for CdWO$_4$.\cite{saatsakis2020temperature} The thermal conductivity values for high entropy (CoCuNiFeZn)WO$_4$ is 2.1 W/m·K at 300\,K. This indicates that the high configurational entropy enhances phonon scattering through local chemical and structural disorder and reduces $\kappa_{ph}$ significantly. Further the decrease in $\kappa_{ph}$ at higher temperature for all the samples may be attributed to the decrease in the mean free path of phonons due to Umklapp process.\\

\begin{figure}
\centering
  \includegraphics[width=0.99\linewidth]{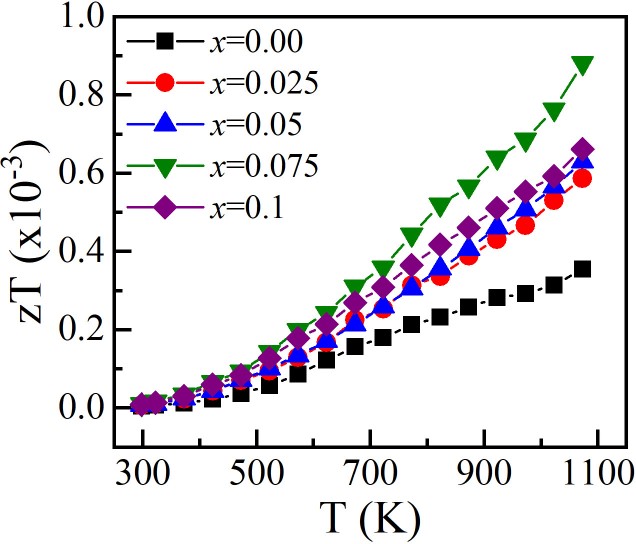}
  \caption{Figure of merit (zT) as a function of temperature for (CoCuNiFeZn)$_{1-x}$Ga$_x$WO$_4$ (0$\leq$\textit{x}$\leq$0.1).}
  \label{fig7a}
\end{figure}
The figure of merit (zT) as a function of temperature calculated from the measured TE parameters for (CoCuNiFeZn)$_{1-x}$Ga$_x$WO$_4$ (0$\leq$\textit{x}$\leq$0.1) is shown in Fig.~\ref{fig7a}. zT increases with rise in temperature for all the samples with a maximum zT of 0.88·10$^{-3}$ at 1100\,K for \textit{x}=0.075.\\
\section{Conclusion}
Ga-doped high entropy wolframite oxides (CoCuNiFeZn)$_{1-x}$Ga$_x$WO$_4$ have been synthesized through solid state reaction followed by spark plasma sintering. The X-ray diffraction pattern indicates single-phase monoclinic structure (space group: \textit{P2/c}) for Ga-doping till \textit{x}=0.075, however with further increase in Ga content, additional WO$_3$ peaks appear. However, SEM-EDS analysis shows the presence of Ga$_2$O$_3$ clusters in equilibrium with the Ga inserted into the wolframite structure with increasing Ga content, which is not seen in the XRD pattern. The cyclic heat treatment of these samples confirms that their thermodynamic stability is enthalpy dominated and hence these are not entropy-stabilized. The optical band gap decreases with Ga doping with a minimum band gap of 1.3 eV for \textit{x}=0.05. The electrical resistivity decreases for small Ga content however the change is not significant for higher \textit{x} values. A proportionate change in the Seebeck coefficient is seen in all the samples, confirming the strong correlation between electrical conductivity and the Seebeck coefficient. Random distribution of several cations at M$^{2+}$ site significantly reduces total thermal conductivity from 11.5 W/m·K for MgWO$_4$ to $\sim$2.1 W/m·K for (CoCuNiFeZn)WO$_4$ which further reduces with Ga doping. As a result of slight improvement in the electrical conductivity and significant reduction in phonon thermal conductivity, we observe an enhanced zT in Ga-doped (CoCuNiFeZn)WO$_4$. However, further doping may be tried for reasonable improvement in zT for these systems.\\
%
%
%
\section*{Conflicts of interest}
There are no conflicts to declare.
\section*{Acknowledgements}
This work was supported by the French Agence Nationale de la Recherche (ANR), through the project NEO (ANR 19-CE30-0030-01). Authors thank the Platform PMCM (Université Paris-Saclay) for the X-ray diffractometer and the "Plateforme de Frittage Ile de France" (Thiais, France) for SPS.\\
\section{Supporting Information}
EDS map for (CoCuNiFeZn)$_{1-x}$Ga$_x$WO$_4$ with \textit{x}=0.025, 0.075. Optical band gap for MgWO$_4$, Thermal conductivity of MgWO$_4$.\\
%


\end{document}